\documentclass[a4paper,english,superscriptaddress,aps,nofootinbib]{revtex4}
\usepackage{amsmath,amssymb,graphicx}
\makeatletter
\usepackage{babel}
\usepackage[active]{srcltx}
\usepackage{graphicx,color}
\usepackage{changebar}
\usepackage{hyperref}
\usepackage[T1]{fontenc}
\usepackage{ae}
\usepackage[ansinew]{inputenc}
\usepackage{amsmath}
\usepackage{braket}
\usepackage{mathtools}
\usepackage{slashed}
\usepackage{empheq}

\usepackage{graphicx}
\usepackage{amsfonts}
\usepackage{amsmath}
\def\x{\xi}

\newcommand{\be}{\begin{equation}}
\newcommand{\ee}{\end{equation}}
\newcommand{\bea}{\begin{eqnarray}}
\newcommand{\eea}{\end{eqnarray}}

\begin{document}

\title{Global Monopole in Palatini $f(\mathcal{R})$ gravity}

\author{J. R. Nascimento}
\email[]{jroberto@fisica.ufpb.br}
\affiliation{Departamento de F\'{\i}sica, Universidade Federal da 
Para\'{\i}ba,\\
 Caixa Postal 5008, 58051-970, Jo\~ao Pessoa, Para\'{\i}ba, Brazil}

\author{Gonzalo J. Olmo}
\email[]{gonzalo.olmo@uv.es}
\affiliation{Departament de F\'{i}sica Te\`{o}rica and IFIC, Centro Mixto Universitat de
Val\`{e}ncia - CSIC,\\
Universitat de Val\`{e}ncia, Burjassot-46100, Val\`{e}ncia, Spain}

\author{P. J. Porf\'{i}rio}\email[]{pporfirio89@gmail.com}
\affiliation{Departament of Physics and Astronomy, University of Pennsylvania, Philadelphia, PA 19104, USA}

\author{A. Yu. Petrov}
\email[]{petrov@fisica.ufpb.br}
\affiliation{Departamento de F\'{\i}sica, Universidade Federal da 
Para\'{\i}ba,\\
 Caixa Postal 5008, 58051-970, Jo\~ao Pessoa, Para\'{\i}ba, Brazil}

\author{A. R. Soares}
\email[]{adriano2da@gmail.com}
\affiliation{Departamento de F\'{\i}sica, Universidade Federal da 
Para\'{\i}ba,\\
 Caixa Postal 5008, 58051-970, Jo\~ao Pessoa, Para\'{\i}ba, Brazil}

\begin{abstract}
	We consider the space-time metric generated by a global monopole in an extension of General Relativity (GR) of the form  $f(\mathcal{R})=\mathcal{R}-\lambda \mathcal{R}^2$. The theory is formulated in the metric-affine (or Palatini) formalism and exact analytical solutions are obtained.  For $\lambda<0$, one finds that the solution has the same characteristics as the Schwarzschild black hole with a monopole charge in Einstein's GR. For $\lambda>0$, instead, the metric is more closely related to the Reissner-Nordstr\"{o}m metric with a monopole charge and, in addition, it possesses  a wormhole-like structure that allows for the geodesic completeness of the space-time.  Our solution recovers the expected limits when $\lambda=0$  and also at the asymptotic far limit. The angular deflection of light in this  spacetime in the weak field regime is also calculated.
\end{abstract}

\maketitle
\section{Introduction}

According to the Big Bang theory, as the universe expanded, its density and temperature decreased resulting in a series of phase transitions which, in association with the mechanism of spontaneous symmetry breaking, may have given rise to several types of topological defects \cite{Kibble}.
Depending on the symmetry group broken, the following defects may have formed: domain walls, cosmic strings, global monopoles, among others. Here we are interested in the Global Monopole (GM), which is described by a field theory with a triplet of coupled scalar fields and presents the following pattern of the spontaneous symmetry breaking: $O(3)\times U(1)$. In short, the vacuum manifold can be shown to be homeomorphic to $S^2\times\mathbb{R}^3$, which is the necessary condition for the formation of this type of defect. In \cite {GM}, Barriola and Vilenkin presented a solution of the Einstein's equations for the metric in the region outside the GM core. They showed that this spacetime has a deficit solid angle and they calculated its impact on the propagation of null geodesics, showing that  light undergoes an angular deflection even if the mass term of the GM is neglected. This effect is quite remarkable because massive particles in the limit of zero GM mass would not feel any Newtonian gravitational pull.  When the GM mass term is not negligible, the solution describes a sort of Schwarzschild black hole with an additional GM charge \cite{Prama}. This kind of geometry is expected  if an ordinary black hole swallows a monopole.

Although General Relativity (RG) has significantly contributed to our understanding of gravitational phenomena, in the last decades we have seen areas such as astrophysics and cosmology raise a series of questions which the theory has difficulties to address in a natural way. In particular,  observations made by two independent groups, Supernova Cosmology Project and Hight-Redshift Supernova Team \cite{SSTC, SCPC,Knop}, together with precise measurements of the temperature fluctuations in the cosmic background radiation and the growth of large scale structures \cite{BoomerangCollaboration,Steinhardt,Halverson}, show that the expansion rate of the universe is positively accelerated, which was previously unexpected from Einstein's equations. On the other hand, theoretical difficulties such as combining GR with the quantum theory or making sense of black hole and cosmic singularities also cast some doubts on the fundamental nature of Einstein's theory.  Faced with these great theoretical challenges, many alternative theories of gravity have arisen (see f.e. \cite{reviews} for some reviews), whose purpose is at least to clarify and guide the research in the description of these phenomena not yet elucidated. Among the variety of different approaches, we find particularly interesting  the so-called Extended Theories of gravity (ETGs) \cite{capozziello2010beyond}, which typically consider modifications of GR based on the addition of higher-order curvature invariants in the usual Einstein-Hilbert action. Such modifications are well motivated by great unification theories, such as string theories.

Once the gravitational sector is extended beyond the GR domain, it becomes necessary to investigate the effects that such modifications could generate in combination with matter theories that are already known. In our case, we are interested in the geometry of spacetime generated by a global monopole. In \cite {barros1997global} Barros and Romero investigated the GM in Brans-Dicke theories using the weak field approximation. In \cite {carames2011gravitational} the same methodology was used but in the context of $f(R)$ theories. The extension of GR used in \cite{carames2011gravitational} motivated a number of other works \cite{carames2012motion, morais2012gravitational, man2013thermodynamic, man2015analytical}. In \cite{carames2017f}, the authors solve exactly the model proposed in \cite {carames2011gravitational} generalizing previous results. In \cite{lambaga2018gravitational}, the authors studied the gravitational field of the GM in the framework of the Eddington-inspired Born-Infeld theory and showed that in the weak-field limit the light deflection is smaller than that obtained by Barriola and Vilenkin in GR.

We would like to emphasize that, with the exception of \cite{lambaga2018gravitational}, all these works were performed in metric formalism without counterpart in the Palatini approach. As is well known, the Palatini formalism is based on the observation that metric and connection are \textit{a priori} equally fundamental and independent geometric quantities \cite{olmo2011}. In the case of Einstein's GR, the connection field equations lead to the Levi-Civita connection as a solution and, therefore, it yields the same predictions as its metric counterpart. However, applying the Palatini formalism beyond GR { shows} that the connection obtained is not  the Levi-Civita one, even if it is symmetric. It is important to emphasize that the field equations generated by the metric formulation are typically of fourth order in derivatives of the metric, while in the Palatini formalism they are second-order. From this perspective, we will discuss the novelties that alternative theories of gravity can bring to the question of the black hole with the GM load in Palatini formalism. Our motivation stems from the work \cite{olmo2016nonsingular} in which the author discussed black hole solutions in { this} formalism in spherically symmetric, static configurations which turn out to be geodesically complete. As we will see, some of our solutions also share this feature. We will also explore the motion of test particles.

 \section{ $f(\mathcal{R})$ THEORY IN PALATINI FORMALISM}
 
 In terms of an arbitrary function of the Ricci scalar, $\mathcal{R}$, the action of the theory reads 
 \begin{equation}\label{acao}
 S=\frac{1}{2\kappa^2}\int d^4x\sqrt{-g}f(\mathcal{R})+S_{\text{mat}}[g_{\mu\nu},\Psi].
 \end{equation}
 Where  $\kappa^2=8\pi G$, and $S_{\text{mat}}[g_{\mu\nu},\Psi]$  is the matter action, which depends on the metric and the matter fields represented by $\Psi$. We { denote} the Ricci scalar { by a calligraphic letter} to highlight that it is constructed out of the contraction of the metric and the Ricci tensor of the connection $\Gamma$: $\mathcal{R}=g^{\mu\nu}R_{\mu\nu}(\Gamma)$. Keep in mind that the Ricci tensor depends on the connection, whose relation to the metric is, {\it a priori}, unknown. 
 The variation of the action with respect to the metric and the connection leads to the following field equations\footnote{These equations are valid also in the case of { having } torsion because those degrees of freedom can be gauged away, see  \cite{Afonso:2017bxr} for details.} \cite{olmo2011} 
 \begin{equation}\label{eq-g}
 f_\mathcal{R}R_{\mu\nu}(\Gamma)-\frac{f(\mathcal{R})}{2}g_{\mu\nu}=\kappa^2T_{\mu\nu}, 
 \end{equation}
 \begin{equation}\label{eq-con}
 \nabla_\lambda(\sqrt{-g}f_\mathcal{R}g^{\mu\nu})=0,
 \end{equation}
 {with} $f_{\mathcal{R}}=\frac{df(\mathcal{R})}{d\mathcal{R}}$. The energy-momentum tensor, which is assumed not to depend on the connection, is given by $T_{\mu\nu}=-\frac{2}{\sqrt{-g}}\frac{\delta S_{\text{mat}}}{\delta g^{\mu\nu}}$. Taking the trace of the equation (\ref{eq-g}), we have
 \begin{equation}\label{traco}
 \mathcal{R}f_\mathcal{R}-2f(\mathcal{R})=\kappa^2T,	
 \end{equation}
 where $T$ is the trace of the  energy-momentum tensor. This relation allows us to write $\mathcal{R}=\mathcal{R}(T)$. Note that if  $T=0$, the Ricci scalar  will be a constant  given in terms of the parameters of $f(\mathcal{R})$. Consequently, $h_{\mu\nu}$ will be proportional to $g_{\mu\nu}$ and therefore, in vacuum, the $f(\mathcal{R})$ theory is reduced to Einstein's GR with a cosmological constant. Then, if we wish to obtain some new physics, we must consider the presence of matter fields in our theory.
 
 If we define an auxiliary metric given by $h_{\mu\nu}=f_{\mathcal{R}}g_{\mu\nu}$, (\ref{eq-con}) implies that the connection associated with $h_{\mu\nu}$ is the Levi-Civita connection: $\Gamma^{\mu}_{\alpha\beta}=\frac{h^{\mu\sigma}}{2}(\partial_\alpha h_{\sigma\beta}+\partial_\beta h_{\sigma\alpha}-\partial_\sigma h_{\alpha\beta})$. Expressing (\ref{eq-g}) in terms of $h_{\mu\nu}$, we are left with
 \begin{eqnarray}\label{eq-h}
 f_{\mathcal{R}}^2R^{\mu}_{\phantom{\mu}\nu}(h)=\frac{\delta^{\mu}_{\phantom{\mu}\nu}}{2}f+\kappa^2T^{\mu}_{\phantom{\mu}\nu}.
 \end{eqnarray}
 Where $R^{\mu}_{\phantom{\mu}\nu}(h)=h^{\mu\alpha}R_{\alpha\nu}(h)$ and $T^{\mu}_{\phantom{\mu}\nu}=g^{\alpha\beta}T_{\alpha\beta}$. We can observe that $R^{\mu}_{\phantom{\mu}\nu}(h)$ corresponds to second order in derivatives of $h_{\mu\nu}$, and the other quantities in (\ref{eq-h}) depend on matter. For a given algebraic structure of the energy-momentum tensor, the equation (\ref{eq-h}) can be solved  to obtain $h_{\mu\nu}$ and, using $h_{\mu\nu}=f_{\mathcal{R}}g_{\mu\nu}$, one obtains $g_{\mu\nu}$ \cite{olmo2016nonsingular}. This is the approach we will follow in the next section. \\
 
A comment on the possibility of dealing with the above equations using a scalar-tensor representation is now in order. In the case of $f(R)$ theories, it is possible to write the action and field equations in such a way that the relevant variables become the metric and a scalar field $\varphi\equiv df/dR$ (see  \cite{Olmo:2005hc,Olmo:2006eh} for details). According to this, one may believe that $f(R)$ theories coupled to a scalar field are equivalent to Einstein's gravity with two (interacting) matter scalar fields. Though this is essentially true in the metric formulation of $f(R)$ theories, it is not so in their Palatini formulation. To see this, it is important to rewrite Eq.(\ref{traco}) in terms of the corresponding scalar field, which leads to 
\begin{equation}\label{eq:ST-pal}
2V-\varphi V_\varphi=\kappa^2T \ ,
\end{equation}
 where $V(var\phi)=R\varphi- f[R(\varphi)]$ and $V_\varphi=dV/d\varphi$. In the metric formulation of $f(R)$ theories, one would get instead 
 \begin{equation}\label{eq:ST-met}
3\Box\varphi+2V-\varphi V_\varphi=\kappa^2T \ .
\end{equation}
 The difference between these two equations is clear. The metric formulation implies a dynamical scalar field, which satisfies the second-order differential equation (\ref{eq:ST-met}), while the Palatini version simply implies an algebraic relation $\varphi=\varphi(T)$. Thus, the way to handle the matter nonlinearities induced by the Palatini case requires new methods and strategies to deal with the field equations. Such methods have been developed and optimized in the last years through the study of more general families of Palatini theories \cite{Makarenko:2014lxa,Olmo:2013gqa,Lobo:2013prg,Olmo:2012nx,Olmo:2009xy}. 
 In this sense, we note that Eq. (\ref{eq-h}) is a particular case of a more general structure that the field equations of Ricci-based gravity theories {\`{a}} la Palatini possess. In general, the field equations of such theories can be written as \cite{Afonso:2017bxr}
  \begin{equation}\label{eq-Omega}
 |\hat\Omega|^{1/2}{R^\mu}_{\nu}(h)=\kappa^2\left({\delta^\mu}_\nu \mathcal{L}_G+{T^\mu}_{\nu}\right),
 \end{equation}
 where $\mathcal{L}_G$ is the gravity Lagrangian and $|\hat\Omega|$ denotes the determinant of a matrix $\hat\Omega$ relating the space-time metric with the metric which solves the connection equation, namely, $h_{\mu\nu}=g_{\mu\alpha}{\Omega^\alpha}_\nu$. In the $f(R)$ case, where $\mathcal{L}_G=f/2\kappa^2$, one finds that  ${\Omega^\alpha}_\nu=f_\mathcal{R}{\delta^\mu}_\nu$ represents a conformal relation between the metrics. In the  general case $h_{\mu\nu}=g_{\mu\alpha}{\Omega^\alpha}_\nu$, however, the relation will not be conformal, which makes the scalar-tensor representation of no use. For this reason we avoid that representation in our analysis.

 \section{GLOBAL MONOPOLE}
 
 The lagrangian density that describes the global monopole is given by
 \begin{equation}
 \mathcal{L}=-\frac{1}{2}(\partial_{\mu}\Psi^a)(\partial^\mu\Psi^a)-\frac{\chi }{4}(\Psi^a\Psi^a-\eta^2)^2.
 \end{equation}
 Where the index $a$, which labels the scalar fields $\Psi^a$, runs from 1 to 3. This model presents spontaneous symmetry breaking $O(3)$ in $U(1)$. The quantities $\chi$ and $\eta$ are, respectively, the coupling constant and the energy scale in which the symmetry is broken. Using the ansatz $\Psi^a=j(r)\eta\frac{x^a}{r}$ and a static, spherically symmetric line element, in \cite{GM} it was found that in the region outside the GM core, where the function\footnote{When the Lagrangian of the matter is replaced by the Lagrangian of the non-linear Sigma model this approximation becomes exact \cite{Prasetyo,Hongwei,Handhika}.} $j(r)\to1$, the energy-momentum tensor is given by 
 \begin{equation}\label{mattertensor}
 T^{\mu}_{\phantom{\mu}\nu}=\text{diag}\left[-\frac{\eta^2}{r^2}, -\frac{\eta^2}{r^2},0,0\right].
 \end{equation}
The fact that this stress-energy tensor has a nonzero trace implies that we will have nontrivial effects in the Palatini formulation of $f(R)$ theories. 

\begin{figure}[h]
 	\centering
 	\includegraphics[height=5.5cm]{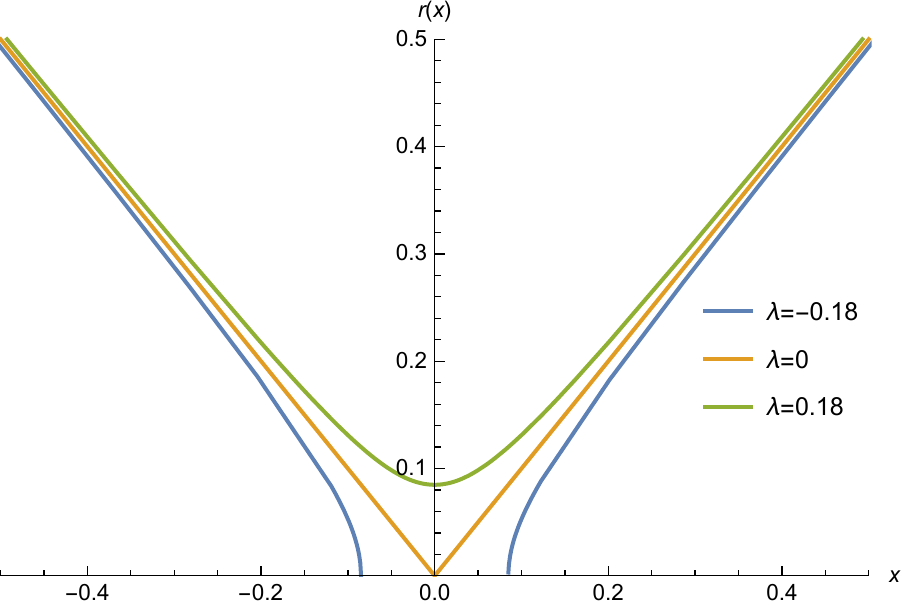}
 	\caption{Function $r(x)$ given by (\ref{z-x}). We are taking $r_0=0.1$}
 	\label{zz}
 \end{figure}

 For static and spherically symmetric space-times, we can adopt the following ansatz for the line element of $h_{\mu\nu}$:
 \begin{equation}
 d\tilde{s}^2=-A(x)e^{2\Phi(x)}dt^2+\frac{dx^2}{A(x)}+\tilde{r}^2(x)(d\theta^2+\sin^2\theta d\phi^2).
 \end{equation}
 Computing $R^{\mu}_{\phantom{\mu}\nu}(h)$, we can show that
 \begin{equation}\label{Rtx}
 R^{t}_{\phantom{t}t}(h)-R^{x}_{\phantom{x}x}(h)=\frac{2}{\tilde{r}}\left(\frac{d^2\tilde{r}}{dx^2}-\frac{d\Phi}{dx}\frac{d\tilde{r}}{dx}\right),
 \end{equation}
 \begin{equation}
 R^{\theta}_{\phantom{\theta}\theta}(h)=\frac{1}{\tilde{r}^2}\left[1-\tilde{r}\frac{d\tilde{r}}{dx}\left(A\frac{d\Phi}{dx}+\frac{dA}{dx}\right)-A\left(\tilde{r}\frac{d^2\tilde{r}}{dx^2}+\left(\frac{d\tilde{r}}{dx}\right)^2\right)\right].
 \end{equation}
 Since $T^{t}_{\phantom{t}t}=T^{x}_{\phantom{x}x}$, we conclude, from equation (\ref{eq-h}), that $\left(\frac{d^2\tilde{r}}{dx^2}-\frac{d\Phi}{dx}\frac{d\tilde{r}}{dx}\right)=0$. Without loss of generality, this result allows us to establish $\Phi(x)=0$ and $\tilde{r}=x$. So we can write the line element
 \begin{equation}\label{metricah}
 d\tilde{s}^2=-A(x)dt^2+\frac{dx^2}{A(x)}+x^2(d\theta^2+\sin^2\theta d\phi^2).
 \end{equation}
 The line element  of $g_{\mu\nu}$ is, therefore,
 \begin{equation}\label{metrica}
 ds^2=-\frac{A(x)}{f_{\mathcal{R}}}dt^2+\frac{dx^2}{f_{\mathcal{R}}A(x)}+r^2(x)(d\theta^2+\sin^2\theta d\phi^2) \ ,
 \end{equation}
 where $r^2(x)$ is related to $x$ via 
\begin{equation} 
x^2=f_{\mathcal{R}}r^2 \ . 
 \end{equation}
 This expression is clearly restricting the domain of the solutions to the region where $f_{\mathcal{R}}$ is positive or zero.  Concerning $R^{\theta}_{\phantom{\theta}\theta}(h)$, its structure, 
 \begin{equation}\label{teta}
 R^{\theta}_{\phantom{\theta}\theta}(h)=\frac{1}{x^2}\left(1-A-x\frac{dA}{dx}\right).
 \end{equation}
 motivates the Ansatze
 \begin{equation}\label{ansatzparaA}
 A(x)=1-\frac{2M(x)}{x}.
 \end{equation}
 From the right-hand side of Eq. (\ref{eq-h}), we find that (\ref{teta}) leads to 
 \begin{equation}\label{dM}
 \frac{2}{x^2}\frac{dM}{dx}=\frac{f}{2f_\mathcal{R}^2}.
 \end{equation}
 To proceed further, and for {the sake of} concreteness, let us consider the model $f(\mathcal{R})=\mathcal{R}-\lambda\mathcal{R}^2$ \cite{staro}, where the constant $\lambda$ has dimensions of length squared. This model implies  $f_{\mathcal{R}}=1-2\lambda\mathcal{R}$, and using the trace (\ref{traco}), we will have $\mathcal{R}=-\kappa^2T$ (exactly like in GR). From (\ref{mattertensor}) we have $\mathcal{R}=\frac{2\alpha^2}{r^2}$, where $\alpha=\kappa\eta$. With this, we can write 
 \begin{equation}\label{fr}
 f_{\mathcal{R}}=1-\frac{4\lambda\alpha^2}{r^2}.
 \end{equation}
  The relation $x^2=f_{\mathcal{R}}r^2$ allows us to write
 \begin{equation}\label{z-x}
 r^2=x^2+4\lambda\alpha^2
 \end{equation}
The Fig. \ref{zz} portrays the behavior of $r$ for different values of $\lambda$ which illustrate the two possible generic configurations and the transient case. Notice that, for $\lambda>0$, $r$ has a  minimum value given by $r=r_{\text{min}}=2\alpha\sqrt{\lambda}$. A minimum value for $r$ implies a solution with a minimum area, just as in a wormhole space-time, with $r_{\text{min}}$ corresponding to the ``wormhole throat''. Later, we will see that this enables a geodesically complete spacetime. For the case $\lambda<0$,  the variable $r$  goes from $0$ up to $\infty$, as we can see from Fig. \ref{zz}. In this case, we have no structure which can lead to wormhole-like solutions and as we will see, this will lead to a geodesically incomplete spacetime.

 Having specified the $f(\mathcal{R})$ model, we have all the necessary elements to solve  Eq.(\ref{dM}), which in terms of $x$, becomes
 \begin{equation}
 	\frac{dM}{dx}=\frac{\alpha^2}{2}+\frac{\lambda\alpha^4}{x^2}
 \end{equation}
 By direct integrating, we are left with
 \begin{equation}
 	M(x)=\frac{\alpha^2x}{2}-\frac{2\lambda\alpha^4}{x}+C,
 \end{equation}
where $C$ is an integration constant which can be chosen as $C=GM$ to recover the GR solution (here $M$ represents the monopole mass and $G$ is Newton's {gravitational} constant). We thus have
\begin{equation}\label{A}
A(x)=1-\alpha^2-\frac{2GM}{x}+\frac{2\lambda\alpha^4}{x^2} \ , 
\end{equation}
and the physical line element (\ref{metrica}) for the GM becomes
\begin{equation}\label{eq:conformal2RN}
	ds^2=\frac{x^2+4\lambda\alpha^2}{x^2}\left[-\left(1-\alpha^2-\frac{2GM}{x}+\frac{2\lambda\alpha^4}{x^2}\right)dt^2+\left(1-\alpha^2-\frac{2GM}{x}+\frac{2\lambda\alpha^4}{x^2}\right)^{-1}dx^2+x^2(d\theta^2+\sin^2\theta d\phi^2)\right].
\end{equation}
Note that this metric is conformally related to a Reissner-Nordstr\"{o}m  metric with electric charge $Q^2=2\lambda\alpha^4$  plus a monopole charge \cite{K}. 
We can see that this solution reduces to the standard GM of GR only when $\lambda=0$ (where $f_{\mathcal{R}}\to1$ and $x=r$) and  in the asymptotic limit $x\to \infty$. Therefore, the effects resulting from the change in Einstein-Hilbert action are appreciable only in the region close to the center of the object. 

\begin{figure}[h]
 	\centering
 	\includegraphics[height=5.5cm]{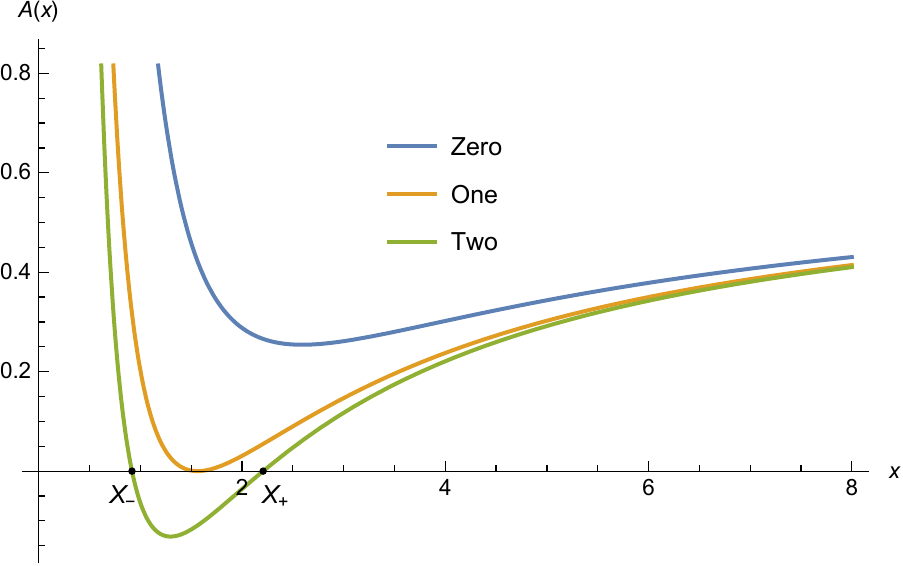}
 	\caption{Function $A(x)$; Zeros correspond to the location of an event horizon.}
 	\label{figA}
 \end{figure}

\begin{figure}[h]
 	\centering
 	\includegraphics[height=6.5cm]{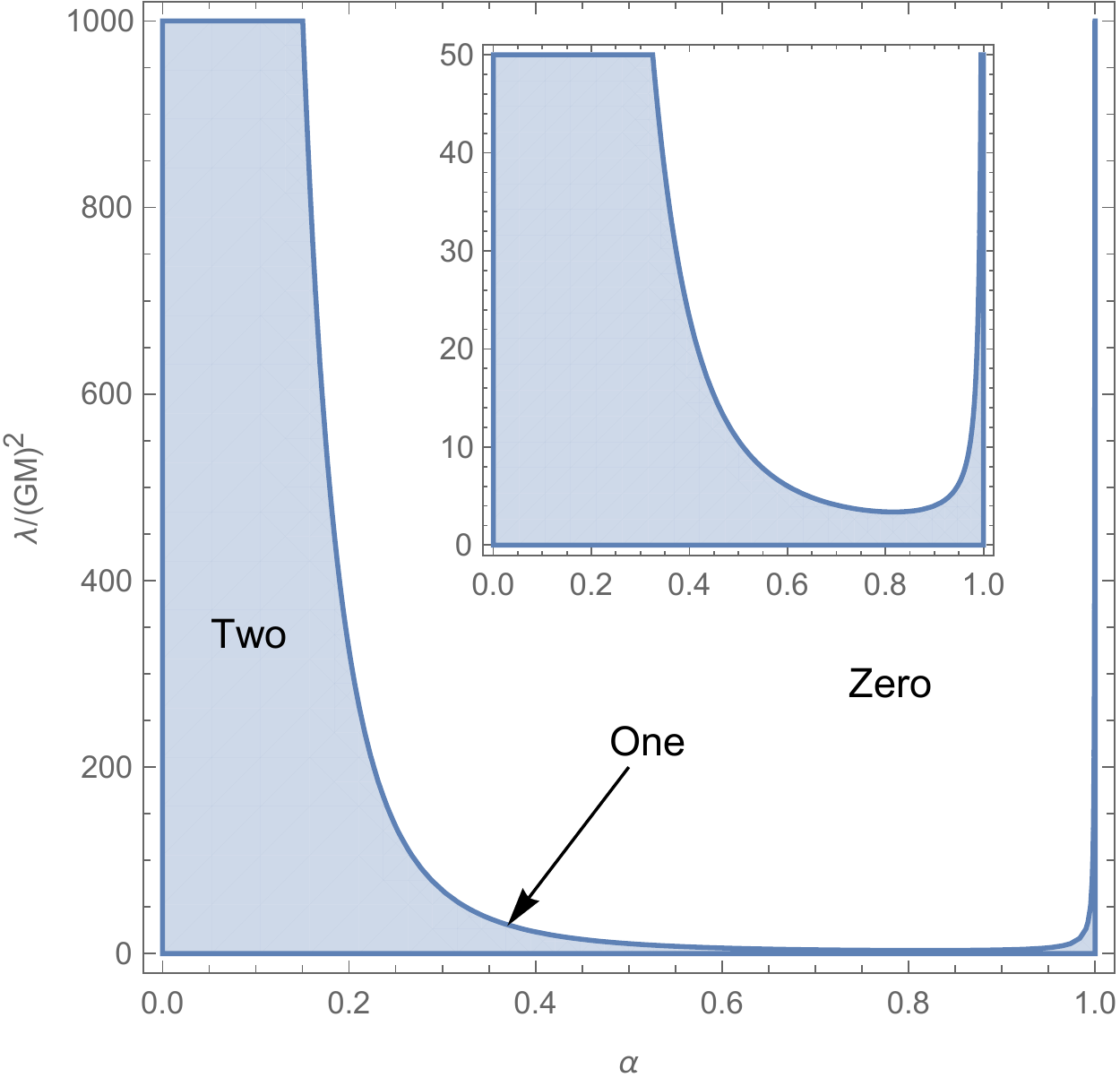}
 	\caption{Regions in which the parameters must be for the solution to satisfy the conditions Zero, One and Two. The smaller graph is only a reduced scale.}
 	\label{restr}
 \end{figure}

The event horizons of this solution appear when $A(x)=0$ and are located at
\begin{equation}\label{hor}
	x_{\pm}=\frac{GM\pm\sqrt{(GM)^2-2\alpha^4\lambda(1-\alpha^2)}}{1-\alpha^2}.
\end{equation}
Note that for $\lambda<0$ the $x$ coordinate is restricted to the domain $x>2\alpha \lambda^{1/2}$ and, therefore, only the $x_+$  horizon is possible.  For $\lambda>0$ there are no restrictions in $x$ and we can find up to two horizons, like in the Reissner-Nordstr\"{o}m black holes, depending on relative values of $GM$ and $2\alpha^4\lambda(1-\alpha^2)$: 
 \begin{itemize}
 	\item No horizons: $(GM)^2<2\alpha^4\lambda(1-\alpha^2)$,
 	\item Degenerate: $(GM)^2=2\alpha^4\lambda(1-\alpha^2)$,
 	\item Two horizons : $(GM)^2>2\alpha^4\lambda(1-\alpha^2)$.
 \end{itemize}
In Fig. \ref{figA} the function $A(x)$ is plotted for the three cases of interest just mentioned.
 In Fig. \ref{restr}, we plot the range of the possible values of the dimensionless quantities $\lambda/(GM)^2$ and $\alpha$ for each of the situations presented above.
 We note that for a realistic value of  $\lambda$ (of the order of a few cm$^2$) and a given value of $\alpha$, say $\alpha=0.2$, for a corresponding large Schwarzschild mass $M$, we will have two horizons (blue region of the figure). As $M$ decreases, in some extreme value, we will have only one horizon (frontier separating the regions and indicated by an arrow in the Fig. \ref{restr}). As $M$ is further decreased, we will arrive in the white region, where $A(x)$ is completely positive, implying the absence of event horizons. In that case one speaks  about a naked singularity. 
  
 \section{GEODESICS} 
  Now that we have {obtained} the metric, the next {step} is to investigate its properties. We will do this through the geodesics associated with this spacetime. Due to the spherical symmetry, it is sufficient to focus on the radial geodesics. In the following sections we will show in which situations the spacetime is geodesically complete. We will consider the effects of the modified geometry on the circular orbits of test particles. 
  
We start by considering the Lagrangian \cite{d1899introducing,wald2010general}
  \begin{equation}
  \epsilon=g_{\mu\nu}\dot{x}^{\mu}\dot{x}^{\nu};
  \end{equation} 
  with $\epsilon=-1,0,1$ corresponding to time-like, null, and space-like geodesics, respectively.
  The dot here denotes the derivative with respect to the affine parameter. Considering the equatorial plane $\theta=\frac{\pi}{2}$, the metric (\ref{metrica}) provides, without loss of generality,
  \begin{equation}\label{lagrangianadageodesica}
  \epsilon=-\frac{A}{f_{\mathcal{R}}}\dot{t}^2+\frac{\dot{x}^2}{Af_{\mathcal{R}}}+r^2(x)\dot{\phi}^2.
  \end{equation}
  The Euler-Lagrange equation for the coordinates $t$ and $\phi$ provide the following conserved quantities:
  \begin{equation}\label{quantidadesconservadas}
  E=\frac{A}{f_{\mathcal{R}}}\dot{t}\quad\text{and}\quad L=r^2(x)\dot{\phi}.
  \end{equation}
  In terms of these new quantities, (\ref{lagrangianadageodesica}) can be written as
  \begin{equation}\label{equacaodeX}
  \frac{\dot{x}^2}{2}=\frac{f_{\mathcal{R}}^2E^2}{2}-\frac{f_{\mathcal{R}}A}{2}\left(\frac{L^2}{r^2(x)}-\epsilon\right).
  \end{equation}
  See that this corresponds to the equation of motion of a classical particle of unit mass, total energy  $\mathcal{E}=\frac{f_{\mathcal{R}}^2E^2}{2}$, subject to the one-dimensional potential $ V_{\text{eff}}=\frac{f_{\mathcal{R}}A}{2}\left(\frac{L^2}{r^2(x)}-\epsilon\right)$. We can thus write 
  \begin{equation}\label{unimotion}
  \frac{\dot{x}^2}{2}=\mathcal{E}-V_{\text{eff}} \ ,
  \end{equation}
where the effective potential takes the explicit form
  \begin{eqnarray}
  V_{\text{eff}}=\frac{x^2}{2(x^2+4\lambda\alpha^2)}\left(1-\alpha^2-\frac{2GM}{x}+\frac{2\lambda\alpha^4}{x^2}\right)\left(\frac{L^2}{x^2+4\lambda\alpha^2}-\epsilon\right).
  \end{eqnarray}
We will next use the above expressions to study the behavior of geodesics in the innermost regions of these solutions, where they depart from the standard GR solution. 

   \subsection{ Massive test particle around the monopole}
   
   As is well known, radial motion will occur whenever $\mathcal{E}-V_{\text{eff}}>0$, and turning points arise when  $\mathcal{E}=V_{\text{eff}}$. Circular orbits of constant radius may occur in regions where the potential is flat, that is, $\frac{dV_{\text{eff}}}{dx}=0$. In this case, we can discriminate two situations: 1)  when $\frac{d^2V_{\text{eff}}}{dx^2}>0$ the orbits will be stable (stable circular orbits or SCO for short), that is, after being slightly displaced from their orbit, particles return rapidly to their original position, and 2) when $\frac{d^2V_{\text{eff}}}{dx^2}<0$, the orbits will be unstable (UCO) and particles will not return to their original orbits after experiencing small perturbations.
   
\begin{figure}[h]
   	\centering
   	\includegraphics[height=5.5cm]{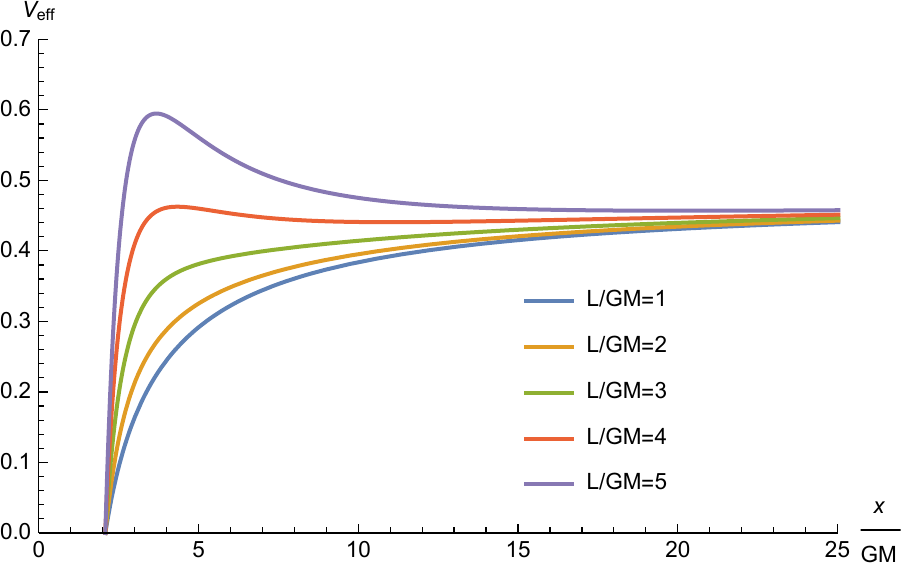}
   	\caption{Effective potential for various angular momentum values. We take $\alpha=0.2$ and  $\frac{\lambda}{(GM)^2}=-0.18$.} 
   	\label{Uf1}
   \end{figure}

\begin{figure}[h]
   	\centering
   	\includegraphics[height=5.5cm]{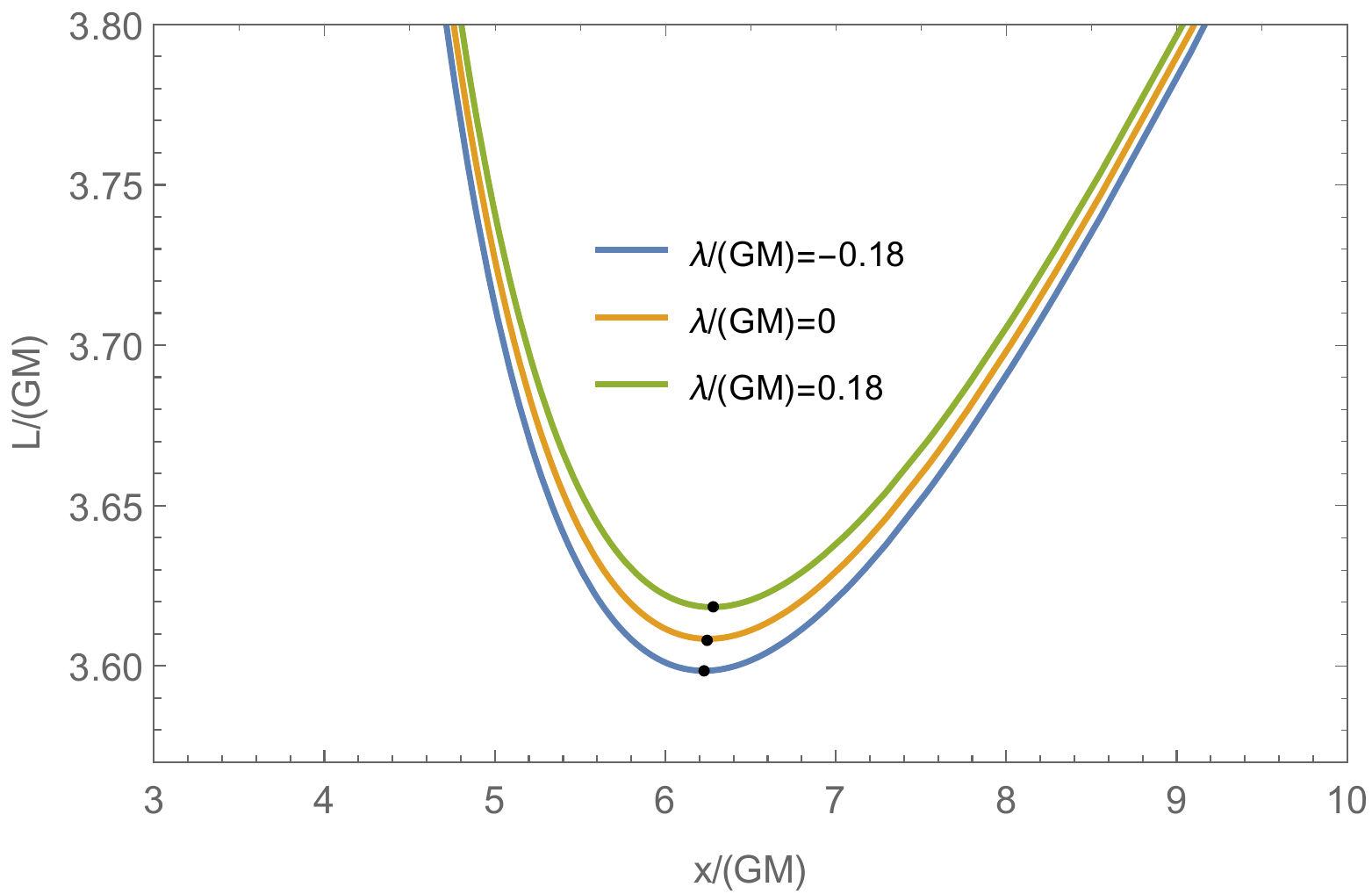}
   	\caption{Circular orbit for several values of $\frac{L}{GM}$. We get it from the equation $\frac{dV_{\text{eff}}}{dx}=0$, solved for $\frac{L}{GM}$.  The black dot indicates the ISCO coordinates. Left side corresponds to unstable orbits and right side, stable.} 
   	\label{inemost-l}
\end{figure}

   Except for null geodesics with $L=0$, for which the effective potential is identically null, the zeros of the effective potential are given by (\ref{hor}), coinciding with the location of the horizons. For $x$ greater than the outer horizon,  the behavior of the potential with respect to the values of $L$ is similar in both cases. Therefore, we will initially consider the effective potential for different types of orbits described by time-like geodesics with ${\lambda}<0$ and for different values of $\frac{L}{GM}$, as shown in the Fig. \ref{Uf1}. We can observe that, when $\frac{L}{GM}$ decreases, the maximum of the effective potential also decreases, while the minimum moves to the left. As we further decrease the value of $\frac{L}{GM}$, the two circular orbits approach until they coincide, producing the innermost stable circular orbit (ISCO), where  $x=x_{\text{ISCO}}$ (see Fig. \ref{inemost-l}). For $x< x_{\text{ISCO}}$, the particle collapses into the black hole. 
   
\begin{figure}[h]
   	\centering
   	\includegraphics[height=5.5cm]{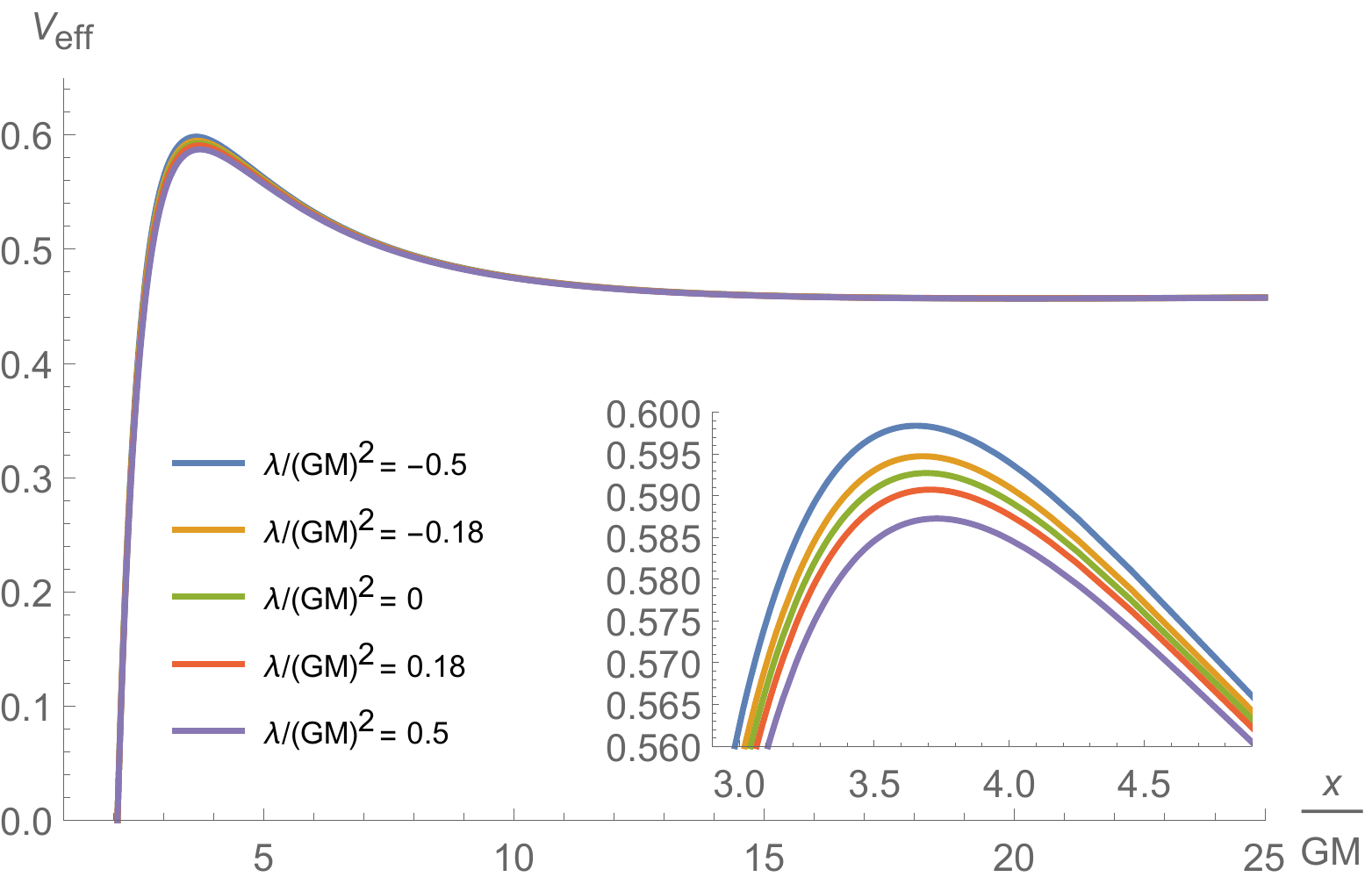}
   	\caption{Effective potential for several values of $\frac{\lambda}{(GM)^2}$. We take $\alpha=0.2$ and $\frac{L}{GM}=5$. Note that $\lambda=0$ corresponds to GR.} 
   	\label{Lambda}
   \end{figure}

   The changes resulting from $\lambda\neq0$ are significant only in the vicinity of the maximum of the potential, as depicted in Fig. \ref{Lambda}. Note that for $\lambda<0$, when $\lambda$ decreases, the maximum of the potential grows and moves away to the left. For $\lambda>0$, when $\lambda$ increases,    the maximum of the potential decreases and moves away to the right. The ISCO changes with respect to the case of GR ($\lambda=0$). When $\lambda<0$, both $L_{\text{ISCO}}$ and $z_{\text{ISCO}}$ decrease. When $\lambda>0$, both $L_{\text{ISCO}}$ and $x_{\text{ISCO}}$ increase, as depicted in Fig. \ref{inemost-l}).
   
\begin{figure}[h]
 	\centering
 	\includegraphics[height=5.5cm]{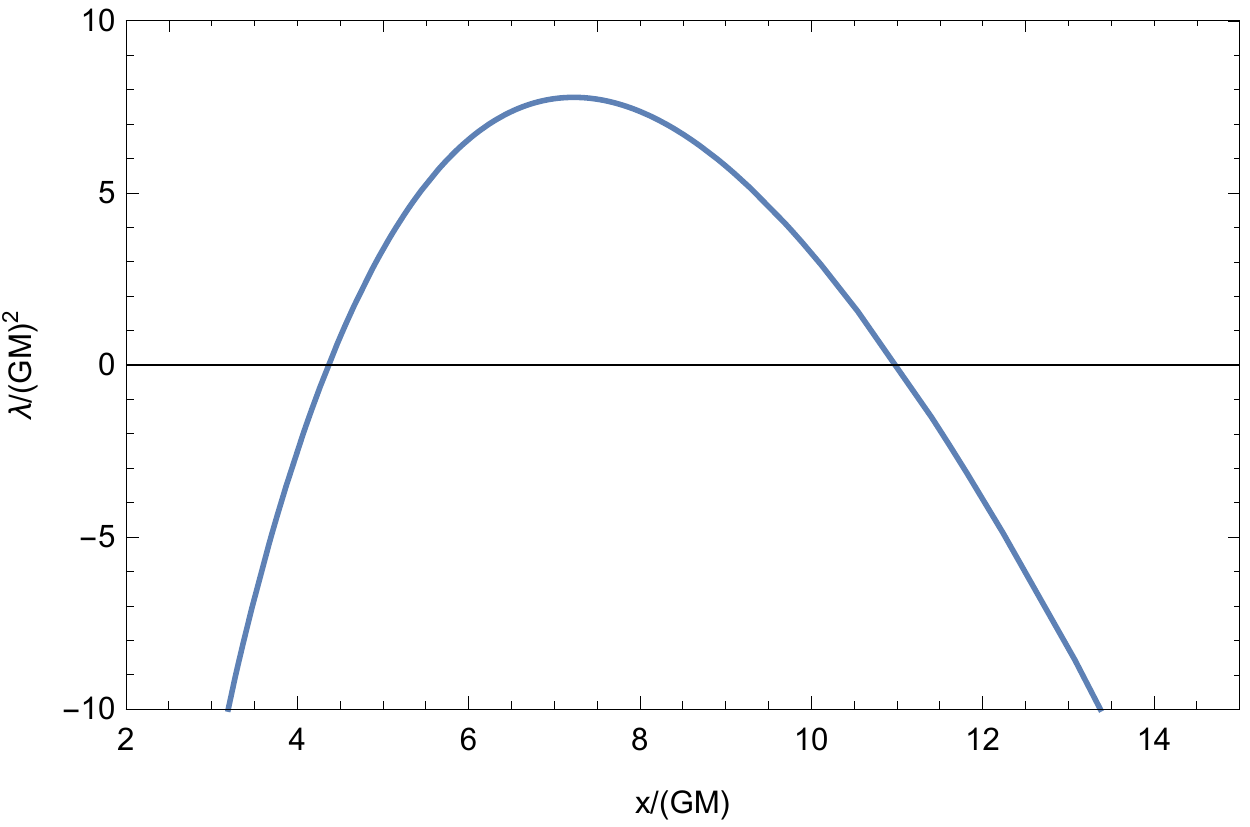}
 	\caption{Circular orbits for several values of $\frac{\lambda}{(GM)^2}$, we take $\alpha=0.2$ and  $\frac{L}{GM}=4$.} 
 	\label{inemost-lamb}
 \end{figure}

Solving $\frac{dV_{\text{eff}}}{dx}=0$ as a function of  $\frac{\lambda}{(GM)^2}$, we observe that above a given value of $\frac{L}{GM}$, there are no circular orbits, as pictured in Fig. \ref{inemost-lamb}.
   
   \subsection{Null geodesics}
   
\begin{figure}[h]
   	\centering
   	\includegraphics[height=5.5cm]{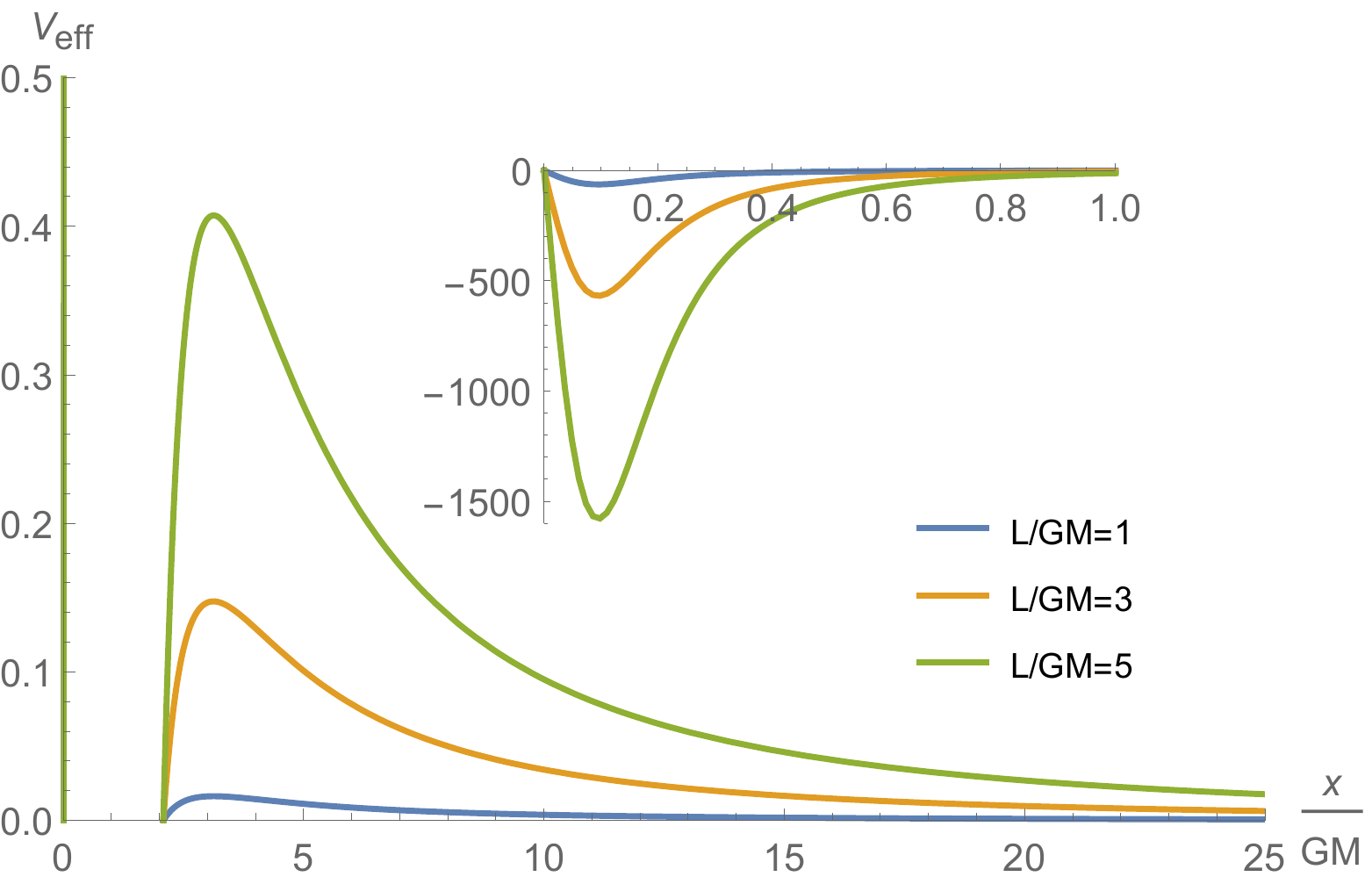}
   	\caption{ Effective potential for null geodesics with $\lambda>0$. The smaller graph illustrates the behavior of the potential near $x=0$, showing the presence of a valley. This is not the Schwarzschild's case, nor  the case of $\lambda>0$ of our model. As always, we are adopting $\alpha=0.2$ e $\frac{\lambda}{(GM)^2}=0.18$} 
   	\label{nula}
   \end{figure}

   Unlike the previous case ($\epsilon=-1$), the shape of the effective potential  for null geodesics ($\epsilon=0$) is independent of $L$. Let us recall that, in the case of the Schwarzschild black hole, the effective potential for $L\neq0$ has a peak corresponding to a single circular orbit. As it is an unstable orbit, any perturbation will make light either fall towards the center of the black hole or be repelled to regions away from the black hole. In our case, the same happens for $\lambda<0$. However, when  $\lambda>0$,  in addition to this peak, the effective potential presents a valley, as illustrated in Fig. \ref{nula}. Although $V_{\text{eff}}$ is flat in the valley, this does not correspond to a stable circular orbit because it lies in between the two horizons.\\ 

   We will now compute  the deflection of light in the spacetime of the black hole with the GM. To do this, we start with the general expression (\ref{equacaodeX}),
   \begin{equation}\label{g1}
   \dot{x}^2=f_{\mathcal{R}}^2E^2-f_{\mathcal{R}}A(x)\frac{L^2}{r^2} \ ,
   \end{equation}
   where the dot denotes the derivative with respect to the affine parameter, and use the formal relation   $x^2=f_{\mathcal{R}}r^2$ to express it in the form 
   \begin{equation}
   \dot{x}^2= f_{\mathcal{R}}^2\left(E^2-A(x)\frac{L^2}{x^2} \right) \ .
   \end{equation}
  From the conserved quantities in Eq. (\ref{quantidadesconservadas}), it is easy to see that $\dot{\phi}=\frac{L}{r^2}$, which allows us to write $\left(\frac{1}{\dot{\phi}}\right)^2=\frac{x^4}{f_{\mathcal{R}}^2L^2}$. Multiplying this expression by the formula above and manipulating, we get to
   \begin{equation}\label{g2}
   \frac{d\phi}{dx}=\left(\frac{x^4}{\beta^2}-x^2A(x)\right)^{-1/2} \ ,
   \end{equation} 
 where we have introduced the apparent impact parameter, $\beta=\frac{L}{E}$. Note that this expression is valid for any $f(R)$ theory and, therefore, can be used to study the light bending in any theory for which an expression for $A(x)$ is known. 
 
 Let $x_0$ be the turning point of an orbit, then from $ \mathcal{E}(x_0)=V_{\text{eff}}(x_0)$ we have 
   \begin{equation}\label{turningpoint}
   \beta^{-2}=\frac{A(x_0)}{x_0^2}
   \end{equation}
   By symmetry arguments, one can see that the contributions to $\Delta\phi$ due to the region before and after the turning point are equal. We can thus write
   \begin{equation}\label{D}
   \Delta\phi=2\int_{x_0}^{\infty}\left(\frac{x^4}{\beta^2}-x^2A(x)\right)^{-1/2}dx.
   \end{equation}
   Now it is convenient to make the change of variable $u=\frac{1}{x}$, in terms of which (\ref{D}) becomes
   \begin{equation}\label{integraldodesvio}
   \Delta\phi=2\int_{0}^{u_0}\left(\frac{1}{\beta^2}-u^2 A(u)\right)^{-1/2}du \ ,	
   \end{equation}
 where $u_0=\frac{1}{x_0}$ and $A(u)=1-\alpha^2-2GMu+2\lambda\alpha^4u$ in our quadratic gravity model. 
 Clearly, the term $2\lambda\alpha^4u^2$  is new as compared to GR. For a typical unification scale, this term is actually very small: if $\alpha^2\equiv \kappa^2\eta^2=8\pi G\eta^2\approx 2.6\times 10^{-5}$ then $\alpha^4\approx10^{-9} $. The integral (\ref{integraldodesvio}) does not have an exact solution. However, we wish to calculate the contribution to the deviation of light only up to the first order in $M$ and $\lambda$. Substituting (\ref{turningpoint}) into (\ref{integraldodesvio}), we are left with
   \begin{equation}\label{integraldodesvio1}
   \Delta\phi=2\int_{0}^{u_0}\left[u_0^2 \left(1-\alpha^2-2GMu_0+ 2\lambda\alpha^4u_0^2\right)  -u^2 \left(1-\alpha^2-2GMu+2\lambda\alpha^4u^2\right)\right]^{-1/2}du.
   \end{equation}
   Expanding $\Delta \phi$ around $\lambda=GM=0$, namely
   \begin{equation}
   \Delta\phi=\Delta\phi\Big|_{GM=\lambda=0}+GM\frac{\partial(\Delta\phi)}{\partial GM}\Big|_{GM=\lambda=0}+\lambda\frac{\partial(\Delta\phi)}{\partial\lambda}\Big|_{GM=\lambda=0}+GM\lambda\frac{\partial^2(\Delta\phi)}{\partial GM\partial \lambda}\Big|_{GM=\lambda=0}+ \mathcal{O}((GM)^2\lambda^2).
   \end{equation}
   We have
   \begin{equation}
   \Delta\phi=\frac{\pi}{\sqrt{1-\kappa^2\eta^2}}+\frac{4GM}{\beta(1-\kappa^2\eta^2)^2}-\frac{3\pi\lambda(\kappa\eta)^4}{(1-\kappa^2\eta^2)^{5/2}}+\frac{6GM\lambda(\kappa\eta)^4}{\beta^3(1-\kappa^2\eta^2)^4}\left(\frac{\pi}{2}+\frac{14}{3}\right)+\mathcal{O}((GM)^2\lambda^2).
   \end{equation}
   The angular deflection up to the first order is, therefore,
   \begin{equation}
   \delta\phi=	\Delta\phi-\pi=\pi\left(\frac{1}{\sqrt{1-\kappa^2\eta^2}}-1\right)+\frac{4GM}{\beta(1-\kappa^2\eta^2)^2}-\frac{\lambda(\kappa\eta)^4}{\beta^3(1-\kappa^2\eta^2)^{5/2}}\left(\frac{3\pi\beta}{2}+\frac{GM(28-3\pi)}{(1-\kappa^2\eta^2)^{3/2}}\right)+\mathcal{O}((GM)^2\lambda^2).
   \end{equation}
   The first term provides the  angular deflection when the  mass term is neglected and, as we can observe, it induces a nonzero light deflection. The second term is the angular deflection for the black hole with mass $M$. Finally, the third term represents the contribution resulting from the modification of the gravitational dynamics generated by the $f(\mathcal{R})$ model we consider. As we said before, this contribution is actually very small, but it can increase or decrease the angular deflection depending on the sign of $\lambda$. For $\lambda>0$, the  correction term is negative, then the deflection is smaller than in the GR case.

  \section{ON THE GEODESIC COMPLETENESS OF SPACE-TIME}
  
  As we have seen, our solution has very distinct characteristics depending on the sign of the constant $\lambda$, which characterizes the extension in the Einstein-Hilbert action. In order to study the completeness of geodesics in this spacetime, we will analyze the behavior of the solution near the center of the object. It should be noted that geodesic completeness is the most standard criterion to determine whether a classical space-time is singular or not \cite{wald2010general,Hawking}. In this regard, the behavior of curvature invariants cannot be regarded as a signature of regularity and should not be used as a shortcut to determine whether a space-time is regular or not \cite{Geroch,Earman,Bejarano:2017fgz,Menchon:2017qed,Olmo:2016fuc,Olmo:2015axa,Olmo:2015bya}. \\
  
  It is well known that in the Schwarzschild solution of GR, the effective potential becomes infinitely attractive within the event horizon and, due to the causal structure, all observers and light rays are conditioned to move in the downwards direction of the radius until reaching the origin, $r=0$. As a consequence, massive particles and light rays falling into the black hole have  incomplete paths because their corresponding affine parameters cannot be extended beyond the center. Following the approach presented in \cite{olmo2016nonsingular}, we will now explore if the space-times resulting from our model are geodesically complete or not. Before going into the details, it is useful to make some observations, since we are working with the  coordinate $x$ instead of the radial function $r(x)$. From equation (\ref{z-x}), when $\lambda<0$, $x$ has a nonzero minimum value ($x_m$) corresponding to $r(x_m)=0$, that is, $x_m=2\alpha\sqrt{|\lambda|}$. On the other hand, when $\lambda>0$, the smallest value of $x$ is zero, which corresponds to the throat of the wormhole ($r=r_{\text{min}}=2\alpha\sqrt{\lambda}$). To start our discussion, it is useful to compute some curvature invariants to see what information they may provide about the innermost regions of our solutions. Focusing on the invariants associated to the metric $g_{\mu\nu}$, whose geodesics describe the motion of test particles and light rays, one finds that both $g^{\mu\nu}R_{\mu\nu}$ and $R^{\mu\nu}R_{\mu\nu}$ diverge as $\x\to 0$ if $\lambda>0$ 
   \begin{eqnarray}
  R(g)&\approx &-\frac{6 \alpha^2}{x^2}+\frac{3M}{\alpha^2\lambda x}+\frac{11}{4\lambda} +\ldots \\
  R^{\mu\nu}(g)R_{\mu\nu}(g)&\approx & \frac{18\alpha^4}{x^4}-\frac{18 M}{\lambda x^3} +\ldots 
  \end{eqnarray}
  and when $x\to x_m$ if $\lambda<0$:
  \begin{eqnarray}
  R(g)&\approx &-\frac{3 \left(3\alpha^3 \sqrt{ \lambda}-2\alpha \sqrt{ \lambda}+2 M\right)}{4(x-x_m)} +\ldots \\
  R^{\mu\nu}(g)R_{\mu\nu}(g)&\approx & \frac{9 \left(9\alpha^6\lambda-12\alpha^4\lambda+4\alpha^2\lambda+4 M^2+12\alpha^3 \sqrt{ \lambda} M-8\alpha \sqrt{ \lambda} M\right)}{16 \left(x-x_m)\right)^6} +\ldots 
  \end{eqnarray}
We thus see that there are curvature divergences for the two types of solutions. If $\lambda>0$ this occurs at the wormhole throat and if $\lambda<0$ it happens at the center.  \\
  
Let  us now consider the extendibility of geodesics starting with $\lambda<0$, for which there is no wormhole-like structure. Inside the event horizon, $V_{\text{eff}}\approx-\frac{L^2x_m}{x^2-x_m^2}$. As in Schwarzschild, the potential becomes infinitely attractive but the effective energy squared diverges more rapidly,  $\mathcal{E}\approx\frac{x_m^4E^2}{(x^2-x_m^2)^2}$, and dominates over the potential so that the geodesics always reach $x=x_m$ in a finite proper time. Since the causal structure prevents particles and light rays from going back in the $x$ direction, the affine parameter freezes at $x=x_m$ with no possibility of extension {\it beyond}. Therefore, this spacetime is geodesically incomplete. To show this more clearly, we will calculate the evolution of the related parameter in the limit $x\to x_m$. From (\ref{unimotion}) we have
  \begin{equation}
  	\frac{dx}{d\tau}=\pm\frac{\sqrt{x_m^4E^2-x_m^2L^2}}{(x^2-x_m^2)},
  \end{equation}
   where $\tau$ is affine parameter and  $\pm$ signs refers to outgoing/ingoing particles (in relation to $x=x_m$). By integrating this expression, we have
  \begin{equation}
  \tau(x)\approx\ \tau_0 \pm\frac{(x_m-x)^2(2x_m+x)}{3\sqrt{x_m^4E^2+x_m^2L^2}}.
  \end{equation}
  where $\tau_0$ is the value of the affine parameter at $x=x_m$, where the $2-$spheres have zero area. Note that when $x\to x_m$, $\tau\to\tau_0$. That is, the affine parameter cannot  go beyond $\tau=\tau_0$, which implies that theses geodesics are incomplete in the past/future. This result is valid for the three types of geodesics (time-like, space-like, and null) and also in the limit $L\to 0$. \\

  In the $\lambda>0$ case, the presence of a wormhole-like structure radically changes the extendibility of geodesics as compared to the previous case. To see this, note that as the minimum at  $x\to0$ is approach, one finds
  $$\lim\limits_{x\to 0}V_{\text{eff}}=\begin{cases}
  \frac{L^2+4\lambda \alpha^2}{16\lambda},&\mbox{if}\quad \epsilon=-1,\\\frac{L^2}{16\lambda},&\mbox{if}\quad\epsilon=0,\\\frac{L^2-4\lambda \alpha^2}{16\lambda},&\mbox{if}\quad \epsilon=1;
  \end{cases}$$
  while $\lim\limits_{x\to 0}\mathcal{E}=0$. Thus, for time-like geodesics the right-hand side of (\ref{unimotion}) must vanish at some point before reaching $x=0$, regardless of the value of $L$. The same happens for null geodesics with $L\neq0$. This means that massive test particles and light rays with nonzero $L$ always encounter a turning point before reaching the wormhole throat at $x=0$.  For null geodesics with $L=0$,  Eq. (\ref{unimotion}) boils down to  
  \begin{equation}
  \frac{dx}{d\tau}\approx\pm E\frac{x}{\sqrt{x^2+4\lambda\alpha^2}}, 
  \end{equation}
  and by direct integration we are left with
  \begin{equation}
  \tau(x)\approx\pm\left[\sqrt{x^2+4\lambda\alpha^2}+2\alpha\sqrt{\lambda}\ln\left(\frac{x}{2\lambda\alpha^2+\sqrt{\lambda\alpha^2(x^2+4\lambda\alpha^2)}}\right)\right].
  \end{equation}
The logarithmic term in this expression puts forward that when $x\to 0$, $\tau\to+\infty$ for incoming rays and $\tau\to-\infty$ for outgoing rays. Thus, the affine parameter is defined over the whole real line, which means that such geodesics are complete. Therefore,  concerning null and time-like geodesics we can already conclude that, for $\lambda>0$, the spacetime is geodesically complete. 

 For the sake of completeness, we now discuss space-like geodesics  ($\epsilon=1$) in the neighborhood  of the wormhole, where $f_{\mathcal{R}}$ rapidly tends to zero. Accordingly, the geodesic equation   can be approximated as
  \begin{equation}\label{eq:spatialgeoapprox}
  \dot{x}\approx \pm\sqrt{\frac{4\alpha^2\lambda-L^2}{16\lambda}-\frac{GM(4\alpha^2\lambda-L^2)x}{16\alpha^4\lambda^2}} \ ,
  \end{equation}
which leads to 
  \begin{equation}
  \tau(x)\approx \tau_0\pm\left(\sqrt{\frac{4\alpha^2\lambda-L^2}{16\lambda}}\right)x.
  \end{equation}
 It is apparent that the affine parameter for these space-like geodesics can be smoothly extended across the throat at $x=0$ to negative values of $x$. These are the only geodesic curves that can effectively cross the wormhole, since time-like ones are repelled before reaching it and radial null geodesics take an infinite time to reach the throat (similar properties can be found for other matter sources \cite{Bambi:2015zch}).  Thus, all types of geodesics are complete in this geometry.
 
Summarizing this section, we have seen that in the $\lambda<0$ case  the spacetime is geodesically incomplete for the three types of geodesics, while in the $\lambda>0$ case, it is geodesically complete. Thus, although both classes of solutions exhibit curvature divergences, only one of them can be regarded as singular from the geodesic perspective.

  \section{Summary and conclusions}
 
We have investigated the space-time geometry generated by a global monopole in a gravity theory with high-curvature corrections of the $f(R)$ type in Palatini formalism. To find the metric, we applied the methodology developed in \cite{olmo2016nonsingular} for block-diagonal stress-energy tensors. The structure of the  energy-momentum tensor corresponding to the exterior of a GM core falls in this category, with the lower block equal to zero, which yields results that differ from other matter sources studied previously \cite{Olmo:2015axa,Bambi:2015zch}.  

For the quadratic model $f(R)=R-\lambda R^2$, we found that the metric is conformally related to a Reissner-Nordstr\"{o}m-like solution with the GM charge, being the black hole charge ($Q^2=2\lambda\alpha^4$) of topological origin [see Eq.(\ref{eq:conformal2RN})]. The conformal relation implies that our solution presents some characteristics which are similar to  the Reissner-Nordstr\"{o}m black hole, such as the location and number of horizons, but there are other aspects which significantly differ. In particular, we have seen that for $\lambda>0$, the space-time has wormhole topology and this plays a key role in the geodesic completeness of the space-time. Aside from this high-curvature behavior, we have verified that in the  asymptotic regions, $|x|\gg1$, our solution reduces to that of Barriola and Vilenkin \cite{GM}. 
  
For $\lambda<0$, the solution is similar to the usual case of GR, though both the Schwarzschild radius and the ISCO radius decrease (only one horizon occurs in this case because the effective charge squared is negative). Still, in this case we have shown that the three types of geodesics are incomplete when $x\to x_m$, (equivalent to $r=0$), implying a singular space-time. This contrasts with the $\lambda>0$ case, for which a minimum area for the $2-$spheres arises giving rise to a wormhole, as mentioned above, and the three types of geodesics are complete. We also verified that the existence of curvature divergences is independent of the geodesic completeness \cite{Bejarano:2017fgz,Menchon:2017qed}. In fact, for $\lambda>0$ curvature scalars blow up at the wormhole throat, $x=0$, while for $\lambda<0$ they explode at the center, $x=x_m$. The former is geodesically complete while the latter is incomplete. 

In the  $\lambda>0$ case, we also found that  the ``wormhole throat" is not a physically accessible region and, therefore, represents a boundary of the space-time. In fact, we have shown that incoming  massive particles and light rays with nonzero angular momentum bounce before getting to the throat, while radial rays take an infinite affine time to get or come out from there.  As a result, there is no need to consider the stress-energy tensor associated to the GM core because the self-consistency of the solutions implies that such core needs not to exist. In other words, a GM of this kind could be supported on topological grounds (wormhole) without additional assumptions on its internal structure (core). This solution could thus be regarded as a geon \cite{geon}. In the $\lambda<0$ case, however, an improved ${T^\mu}_\nu$ could help alleviate the singular character of the geometry.  

In a forthcoming paper we will investigate if the qualitative aspects of the solutions found in the quadratic $f(R)$ model studied here still persist when other curvature invariants, such as $R_{\mu\nu}R^{\mu\nu}$, are taken into account in the action. This is relevant to determine the robustness of our findings.

\textbf{Acknowledgments.}  
P. J. Porf\'{i}rio would like to acknowledge the Brazilian agency CAPES (PDE process number 88881.171759/2018-01) for the financial support. GJO is funded by the Ramon y Cajal contract RYC-2013-13019 (Spain).  The work by A. Yu. P. has been supported by the
CNPq project No. 303783/2015-0.
This work is supported by the Spanish projects FIS2014-57387-C3-1-P  and FIS2017-84440-C2-1-P (MINECO/FEDER, EU), the project H2020-MSCA-RISE-2017 Grant FunFiCO-777740, the project SEJI/2017/042 (Generalitat Valenciana), the Consolider Program CPANPHY-1205388, and the Severo Ochoa grant SEV-2014-0398 (Spain). This study was financed in part by the Coordena\c{c}\~ao de Aperfei\c{c}oamento de Pessoal de N\'{i}vel Superior - Brasil (CAPES) - Finance Code 001.

\end{document}